\documentclass[12pt,leqno]{article}
\usepackage{amsmath,amsfonts,caption2}
\usepackage{graphicx}
\begin{document}

\begin{center}
\textbf{RELATIVISTIC CONTRIBUTIONS TO THE PHOTON\\
       ASYMMETRY IN DEUTERON PHOTODISINTEGRATION}

       \medskip

K.Yu. Kazakov and D.V. Shulga

\medskip \small{ Laboratory of
Theoretical Nuclear Physics,\\ Far Eastern State University,\\
Sukhanov Str. 8, Vladivostok, 690600, Russia.}
\end{center}

\begin{abstract}
The cross section asymmetry $\Sigma$ in deuteron disintegration with
linearly polarized light is considered in the framework of the
Bethe-Salpeter formalism. Relativistic contributions to the asymmetry are
investigated within the approximation, which takes into account the
one-body part of the electromagnetic current operator and the dominant
positive-energy partial amplitudes of the Bethe-Salpeter vertex function
of the deuteron.
\end{abstract}


\section{Introduction}

Interest to the study of beam-polarization observables (namely,
the polarization-difference cross sections and the cross-section
asymmetry) in the exclusive deuteron photodisintegration is
twofold. In the energy range from below pion threshold to above
the $\Delta$ resonance, it is to find features characterizing
sensitivity of these observables to the short-range behavior of
the tensor force in models of the nucleon-nucleon (NN)
interaction~\cite{legs1}. The study of the polarization
observables also offers a testing ground for models of the
$N\Delta$ interaction, as the polarization-dependent cross
sections are sensitive to interference terms due to the
meson-exchange as well as delta-isobar currents
(Ref.~\cite{legs2,mami} and references therein). On the other
hand, recent experiments on two-body photodisintegration of the
deuteron at the TJNAF and the Yerevan Synchrotorn are devoted to
measuring both the differential cross section and the asymmetry in
the range, when the photon energy is commensurate with the rest
mass of the nucleon. Instead of the reaction mechanism related to
the conventional mesonic picture of the NN interaction, here one
looks for the partonic phenomena such as the scaling of the cross
sections and the hadronic helicity
conservation~\cite{bochna,adamian}.

Relativistic contributions (RC) are expected to be significant in
deuteron photodisintegration, especially when the photon momentum
involved in the process, becomes comparable with the rest mass of
the nucleon. So far, a through treatment of RC has been based on
the Blankenbecler-Sugar reduction of the Bethe-Salpeter (BS)
equation with spinor reduction carried out to the order of
$\mathcal O(\tfrac{1}{m^2})$ for the charge and current
densities~\cite{jaus}. In the nonrelativistic formalism
relativistic corrections has been investigated in a number of
papers, for example see Ref.~\cite{ying} where it is shown that
the inclusion of RC leads to an important reduction of the forward
differential cross section. More recently, the strong influence of
relativity is found in analysis of the Gerasimov-Drell-Hearn sum
rule for the deuteron in the photon energy range up to about 100
MeV~\cite{GDH}. Inclusion of leading order RC, this is the
spin-orbit term, in the photodisintegration channel results in
large positive contribution reducing the value of the sum rule in
absolute size by more than 30 percent!

In this talk we present some results concerning the investigation
of the RC in the beam-polarization observables in deuteron
photodisintegration. This is a continuation of the approach of
Ref.~\cite{mypaper} applied to the analysis of the differential
cross section. A relativistic formulation is based on the
Bethe-Salpeter (BS) equation for the description of the dynamics
of 2N system and the gauge invariant procedure for the
construction of the electromagnetic (EM) current.

\section{Definition of observables}

The kinematics of the process
\begin{eqnarray}
\Vec\gamma(q)+D(\mathbb K)\to P(k_p) + N(k_n)
\end{eqnarray}
for disintegration of the deuteron induced by a linearly polarized
photon in the c.m. frame is organized as follows: $\mathbb
K=(\sqrt{M_d^2+\mathbf K^2},\mathbf K)$ and
$q=(\omega,\boldsymbol{\omega})$ is the four momenta of the
deuteron and the real photon, respectively, with $\mathbf K
+\boldsymbol{\omega}=\mathbf 0$; $k_p$ and $k_n$ are the four
momenta of the proton and neutron, $k_{p,n}^2=m^2$. Further, it is
convenient to introduce the total $\mathbb P=k_p+k_n$, $\mathbb
P\equiv(\sqrt{s},\mathbf 0)$, and the relative asymptotic
$p=\tfrac12(k_p-k_n)$, $p=(0,\mathbf k_p)$, four momenta.

The differential cross section for complete linear polarization of the
photon is written in the form
\begin{equation}
d\sigma(E_\gamma,\Theta,\Phi)=
d\sigma_0(E_\gamma,\Theta)[1+\Sigma(E_\gamma,\Theta)\cos(2\Phi)],
\end{equation}
where $d\sigma_0$ is the differential cross section in the case of the
unpolarized initial configuration, $\Sigma$ is the cross section azimuthal
asymmetry, $\Theta$ is the angle between the momentum of the photon
$\boldsymbol\omega$ and that of the outgoing proton $\mathbf k_p$, and
$\Phi$ is the angle between the polarization plane and the reaction plane.
If the photon three momentum is taken to be in the $+z$ direction
$\boldsymbol\omega=(0,0,\omega)$, the latter is defined by the
polarization vector of the photon $\boldsymbol\epsilon_\lambda$ and
$\boldsymbol\omega$ and the former --- by vectors $\boldsymbol\omega$ and
$\mathbf k_p$. In the transverse gauge for the radiation field, following
the property $\boldsymbol\omega\cdot\boldsymbol\epsilon_\lambda=0$, one
may write for the photon helicities $\lambda=\pm 1$
\begin{equation}
\boldsymbol\epsilon_\pm=\mp\tfrac{1}{\sqrt{2}}(0,1,\pm\imath,0).
\end{equation}

The asymmetry $\Sigma$ is the ratio of the difference to the sum
of cross sections obtained with the photon's electric vector
parallel and perpendicular to the reaction plane, respectively
[$\Sigma=(d\sigma_{||}-d\sigma_\perp)/(d\sigma_{||}+d\sigma_\perp)]$.
Both the polarization-independent cross section
$d\sigma_0/d\Omega=
\tfrac12(d\sigma_{||}/d\Omega+d\sigma_\perp/d\Omega)$ and the
polarization-difference cross sections
$d\Delta/d\Omega=\tfrac12(d\sigma_{||}/d\Omega-d\sigma_\perp/d\Omega)$
are expressed in terms of the invariant reduced amplitude
$t_{Sm_s\,\lambda m_d}(\Theta)$, depending on the photon helicity
$\lambda$ and projections of the good total angular momentum of
the deuteron $m_d=\pm1, 0$ and of the total spin ($S=0,1$) of np
pair $m_s$ on to quantization axis
\begin{align}
\dfrac{d\sigma_0}{d\Omega_p} &=\dfrac{\alpha}{16\pi s}\dfrac{|\mathbf
k_p|}{\omega} \sum\limits_{Sm_S\lambda m_d} |t_{Sm_s\,\lambda m_d}|^2,\\
\Sigma&=-\frac{\sum\limits_{Sm_S\lambda m_d} t_{Sm_s\,\lambda
m_d}t_{Sm_s\,-\lambda m_d}^*} {\sum\limits_{Sm_s\lambda
m_d}|t_{Sm_s\,\lambda m_d}|^2}, \label{sigma-T}
\end{align}
where $\alpha=e^2/(4\pi)$ is the fine-structure constant, and taking the
cross section as functions of the laboratory photon energy $E_\gamma$ and
the c.m. angle $\Theta$, one relates the variables as
\begin{equation}
s=M_d(M_d+2E_\gamma),\quad |\mathbf k_p|=\sqrt{\frac{s}{4}-m^2},\quad
\omega=\frac{M_d}{\sqrt{s}}E_\gamma. \label{kin-s}
\end{equation}

\section{Few theory}

Notwithstanding that the nonrelativistic interpretation of
deuteron photodisintegration is successful at low and intermediate
energies, a systematic exposition of the covariant formalism at
this energy range for the 2N system is still missing. Essential
ingredients of the nonrelativistic formulation are the pure
one-body EM plus meson-exchange currents as incorporated in the
Siegert operators, explicit pair and pionic current, the
spin-orbit and further relativistic corrections and delta-isobar
configurations~\cite{arenhovel}. At present it is impossible to
claim quantitative description of the process in question, using
the fully relativistic analysis with the inclusion of the all
relevant physics. In order to study RC here we apply the
four-dimensional (4D) formalism of the BS equation. The reason for
the application of the 4D-formalism is to preserve generality and
to avoid some difficulties arising in considering EM process on
few nucleon systems. It is known the relativistic effects can take
different forms depending upon the organization of the dynamics
and its forms~\cite{wallace}. And, if a quasipotential formalism
is chosen with the same organization of the 2N interaction,
equivalent results should be found in reduction of the exact BS
matrix elements. The leading terms of such reduction should
correspond to the nonrelativistic approximation, and the next
order terms should produce $\mathcal
O(\tfrac{1}{m^n})$-corrections.

In the BS formalism the matrix elements associated with the
physical observables in Eq.~\eqref{sigma-T} are obtained via the
recipe for the EM current operator construction of
Refs.~\cite{shebeko,ito}, which is consistent with the form of
the NN interaction kernel of the BS equation and gauge
invariance. The dynamical model we assume in this work is the
one-body plane-wave approximation described in
Ref.~\cite{mypaper}. In terms of the BS amplitudes for the
deuteron $\psi_{\mathbb K m_d}$ and the np pair $\chi_{\mathbb P
p\,Sm_s}$, the c.m. transition amplitude is written in the form
\begin{equation}\label{t-amp}
T_{Sm_s\,\lambda m_d}(\Theta,\Phi)= \frac{1}{4\pi^3}\int\!\! d^4k
d^4l\: \bar{\chi}_{\sqrt{s}{\bf p}\,Sm_s}(l)\,
\epsilon_\lambda\cdot\Lambda(l,k;\mathbb P,\mathbb
K)\,\psi_{\mathbb K m_d}(k),
\end{equation}
where $T_{Sm_s\,\lambda
m_d}(\Theta,\Phi)=\text{e}^{\imath(\lambda+m_d)\Phi}t_{Sm_s\,\lambda
m_d}(\Theta)$, $\Lambda$ denotes the Mandelstam vertex, which
determines the EM interaction with 2N system in the framework of
the BS formalism. Here all amplitudes are the $4\times4$ matrices
in the spinor subspace. The four momentum conservation at the
photon-deuteron vertex gives $\mathbb P=\mathbb K+q$. Owing to the
fact that the EM interaction does not conserve the total isotopic
spin, the invariant amplitude in Eq.~\eqref{t-amp} contains
isovector ($\Delta T=1$) and isoscalar ($\Delta T=0$) transitions
corresponding to a given spin state $|\mathbb P p\,Sm_s\rangle$ of
the np pair. The complete description of our formalism can be
found in Ref.~\cite{mypaper}.

\section{Analysis of relativistic effects}

There are two kinds of the relativistic corrections to the
nonrelativistic impulse approximation charge and current
densities: (i) terms generated from the spinor reduction of the
EM current operator, and (ii) terms generated from boosting of
the initial deuteron wave function from the rest frame to the
c.m. frame of the reaction with the deuteron moving with a
velocity $\tfrac{\omega}{M_d}$~\cite{jaus,ying}. In a more
complete theoretical description, in addition to the one-body EM
current, meson-exchange currents, giving rise to two-body
operators, are included. Here the inclusion of energy transfer
effects are important~\cite{hwang,schwamb}.

Using the BS formalism one may develop, in principle, systematic
studies of the relativistic effects in the EM interactions with
2N systems. Here one distinguishes several classes of the
relativistic effects both of the kinematical and dynamical
origin. First of all, the relativistic effects in the internal
dynamics brought by the generalized relativistic wave function of
the deuteron in its rest frame. These are effects of the
negative-energy partial-states in the vertex function, the
retardation (the relative-energy dependence), and the nonstatic
single-particle propagators.

Second, one has to boost the wave function to a moving frame as
\begin{equation}
\psi_{\mathbb Km_d}(k)=\Lambda^{(1)}(\mathcal
L)\Lambda^{(2)}(\mathcal L)\psi_{\mathbb K_{(0)} m_d}({\mathcal
L}^{-1}k), \label{boost-vertex}
\end{equation}
where $\Lambda^{(i)}({\mathcal L})$, $i=1,2$ is the booster to the
c.m. frame and $\mathcal L$ is the Lorentz transformation matrix
corresponding to $\mathbb K_{(0)}=(M_d,\mathbf 0)$. The correct
treatment means that the spin booster $\Lambda^{(i)}({\mathcal
L})\neq1$ (the effect of the spin precession) and the effect of
boosting on the internal space-time variables $\mathcal L\neq1$
(the Lorentz contraction).

\begin{figure}[thp]\label{fig1}
\setlength{\abovecaptionskip}{0mm} \setlength{\belowcaptionskip}{5mm}
\captionstyle{centerlast}
\renewcommand{\captionlabeldelim}{.~}
\renewcommand{\captionfont}{\linespread{1.25}\small\sl}
\renewcommand{\captionlabelfont}{\normalsize\sffamily}
\setcaptionwidth{0.8\textwidth}
\onelinecaptionsfalse \caption{Some relativistic contributions to
the cross section asymmetry. Solid line is the exact
positive-energy one-body BS calculation, dot-dashed line is the
static approximation to the exact result, dotted line is the
static approximation with taking into account the booster
$\Lambda$, and dashed line is the nonrelativistic impulse
approximation. The transverse gauge is used for all the curves.}
\centering
\includegraphics[width=0.75\textwidth]{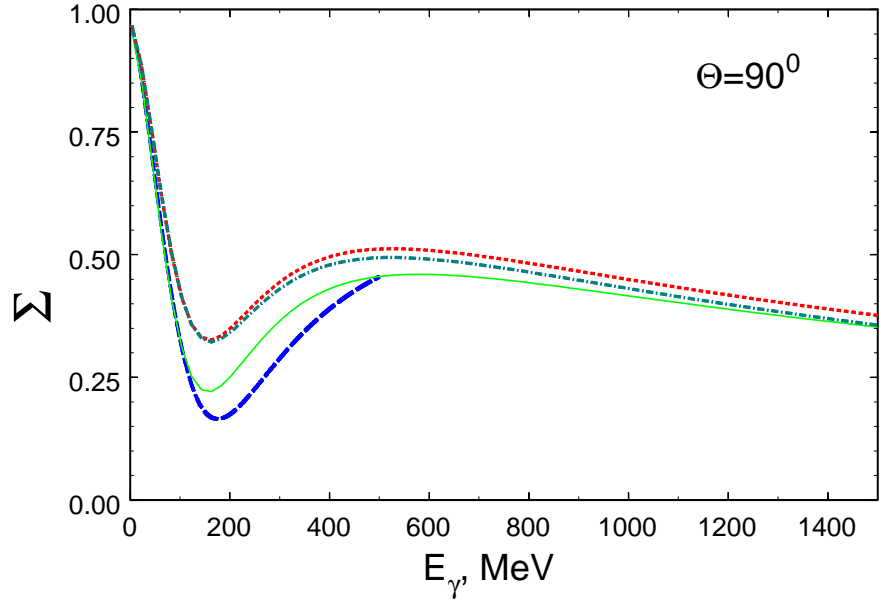}
\end{figure}

Third, the correct treatment of the conserved EM current operator. A
method of minimal substitution can be used  for constructing the EM
two-body current operator of two-nucleon system within the BS
formalism~\cite{ito,shebeko}. The relativistic two-body current operator
is derived using the Ward-Takahashi identities for the one-body current
operators and the relativistic equation for the bound state. The result
depends on the charge-exchange and nonlocal properties of the NN
interactions.
\begin{figure}[ht]\label{fig2}
\setlength{\abovecaptionskip}{0mm} \setlength{\belowcaptionskip}{5mm}
\captionstyle{centerlast}
\renewcommand{\captionlabeldelim}{.~}
\renewcommand{\captionfont}{\linespread{1.25}\small\sl}
\renewcommand{\captionlabelfont}{\normalsize\sffamily}
\setcaptionwidth{0.8\textwidth}
\onelinecaptionsfalse \caption{Dependence of the cross-section
asymmetry on the interaction kernel of the BS equation. Dashed
curve is the positive-energy BS calculation with the
one-boson-exchange vertex function, dot line is the same with
Graz-II vertex function. Shaded area shows the effects of the
retardation in the Graz-II vertex function. } \centering
\includegraphics[width=0.75\textwidth]{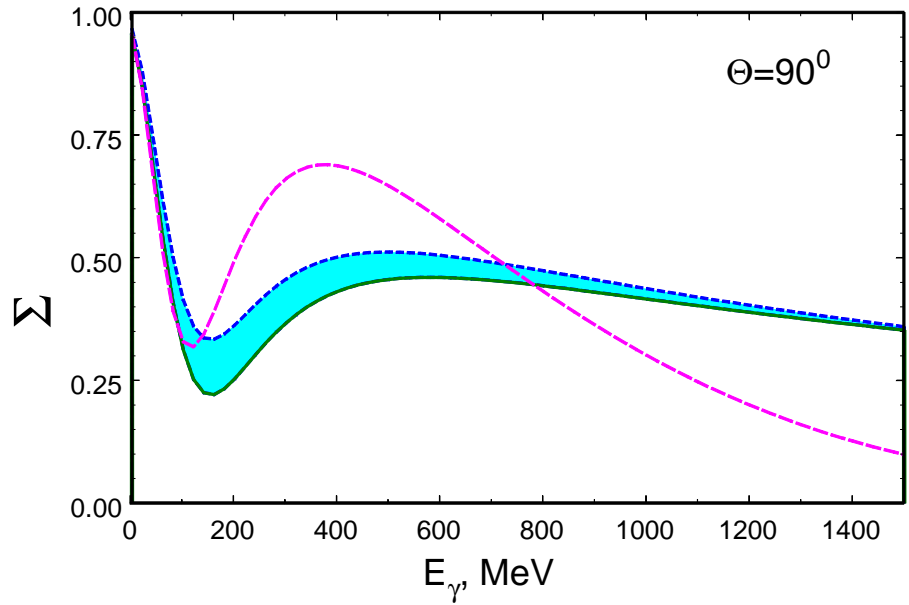}
\end{figure}
The matrix element in Eq.~\eqref{t-amp} is calculated in the
one-body plane wave approximation with the transverse gauge. The
vertex function is the solution of the homogeneous BS equation
with the Graz-II separable interaction~\cite{tjon}. The dominant
positive-energy components of the vertex function of the deuteron
are considered. In Fig.~\ref{fig1} we present results of the
dependence of the asymmetry at $\Theta=90^\circ$ on the
laboratory photon energy $E_\gamma$. Here the curves concerning
the number of approximations with respect to the exact
positive-energy BS calculation  (solid curve) are depicted. The
static approximation (dot-dashed curve) amounts to neglecting the
boost on the internal and spin variables of the BS amplitude for
the deuteron. Such RC as the Lorentz contraction and spin
precession are excluded. The effect due to the booster on the
spin variables is shown by dotted line in Fig.~\ref{fig1}.

We find that the asymmetry $\Sigma$ bears rather strong dependence
on the interaction kernel used in solving the BS equation for the
deuteron. Its is shown in Fig.~\ref{fig2}. The solid and dash
curves are the calculations with the positive-energy vertex
functions of the deuteron, which are solutions of the BS equation
with the one-boson-exchange and Graz-II interaction (strength of
the D-state is equal to 5\%) kernels, respectively. Since a
realistic BS amplitude is computed using the Wick
rotation~\cite{umnikov}, it can be taken in our analysis only at
the relative energy equal to zero. The effect of the retardation
(the relative energy dependence of the vertex function) in case of
the separable interaction is highlighted in Fig.~\ref{fig2}. The
deviations between two curves are explained by, first, that the
minima in the S-waves of the two vertex functions are shifted on
approximately 100 MeV, and, second, D-wave of the realistic vertex
function is much softer than that of the separable one. The latter
is rather crucial for the behavior of the asymmetry for the photon
energies higher than 600 MeV.

In summary, we conclude that the relativistic effects could make a
significant contribution to the polarization-beam observables in
deuteron photodisintegration, though for their simple assessment
one should encourage us to conduct a more complete analysis.

\end{document}